\documentclass[11pt]{article}
\usepackage{epsfig, amssymb}
\usepackage{amsmath}
\usepackage{caption}
\usepackage{subcaption}
\begin{document}
\newcommand{\pa}{\partial}
\newcommand{\e}{\varepsilon}
\newcommand{\beq}{\begin{equation}}
\newcommand{\eeq}{\end{equation}}
\newcommand{\bef}{\begin{figure}}
\newcommand{\eef}{\end{figure}}
\title{
Spin Wave Neuroanalog of von Neumann's Microwave Computer
}
\author{Frank Hoppensteadt\footnote{
This work was presented at Neural Coding 2014, Versailles, France, October 6, 2014. The author is supported in part by Neurocirc LLC. \textbf{Keywords:} Neuromorphic computing. Spin torque oscillator. Spin waves. Logic machine. Nanomagnetics. von Neumann microwave computer.
}\\
fh21\_at\_nyu\_dot\_edu\\
Courant Institute/NYU\\
Neurocirc LLC.}
\date{}
\maketitle
\begin{abstract}
Frequency and phase of neural activity play important roles in the behaving brain. The emerging understanding of these roles has been informed by the design of analog devices that have been important to neuroscience, among them the neuroanalog computer developed by O. Schmitt in the 1930's. In the 1950's, J. von Neumann, in a search for high performance computing using microwaves, invented a logic machine based on similar devices, that can perform logic functions including binary arithmetic.  Described here is a novel embodiment of his machine using nano-magnetics. The embodiment is based on properties of ferromagnetic thin films that are governed by a nonlinear Schr\"odinger equation for magnetization in a film. Electrical currents through point contacts on a film create spin torque nano oscillators (STNO) that define the oscillator elements of the system. These oscillators may communicate through directed graphs of electrical connections or by radiation in the form of spin waves. It is shown here how to construct a logic machine using STNO, that this machine can perform several computations simultaneously using multiplexing of inputs, that this system can evaluate iterated logic functions, and that spin waves can communicate frequency, phase and binary information. Neural tissue and the Schmitt, von Neumann and STNO devices share a common bifurcation structure, although these systems operate on vastly different space and time scales. This suggests that neural circuits may be capable of computational functionalities described here.

%submitted NC2014abstract.rtf 6/17/14
\end{abstract}

Various aspects of frequency and phase in neuroscience have been widely investigated (e.g., see \cite{desai,nunez,EMI2007}, for further references). Models for observed or speculated neural phenomena have often been based on nonlinear oscillators that describe electromagnetic, biochemical or mechanical processes. Emerging knowledge of neural dynamics has influenced, and has been influenced by such models. In particular, there have been close ties between emerging knowledge of neural processes and the development of modern computers. 

In his studies of action potentials in squid axons in the 1930's, O. Schmitt designed a device, now known as the Schmitt trigger, which he used to build a neuroanalog computer for his studies. von Neumann used similar devices in the 1950's in his invention of a logic machine based on microwave oscillators. Neural tissue and the Schmitt, von Neumann and STNO (spin torque nano oscillator) devices described here share a common Andronov-Hopf (AH) bifurcation structure, although these systems operate on vastly different space and time scales. From a mathematical point of view, these devices are closely related to the Hodgkin-Huxley model and various heuristics for it \cite{HHH}. 

Section 1 describes von Neumann's invention \cite{JvN} for performing binary arithmetic using oscillator systems. While he did not give specific designs, he did provide the essential elements for constructing a functioning machine. Others followed with specific designs using oscillators ranging from cryogenic Josephson junctions to various phase locked loops and relaxation oscillators. 

Section 2 reveals a novel way in which von Neumann's invention can be embodied in spin torque nano oscillators (STNO). The STNO model here has been derived and studied in \cite{HKM2011, astno}, and this article expands the presentation \cite{NC2014}. Other applications of this methodology, not shown here, are to switching and control. Examples show how a single STNO can perform simple logic computations, how an array of STNO can evaluate iterated logic functions, and how a single STNO can perform several separate computations simultaneously by multiplexing of inputs. Section 3 shows how spin waves can transmit and process binary information using phase and frequency. 

This article is about physical devices and is not about the brain. It is not clear how the results here might be interpreted in the context of the brain, nor how magnetic behavior in the brain might occur or what its functions might be, although there are magnetic elements in the brain, so spin waves or their kin may be present.  The novelty here is in showing that any logic computation may be performed by spin torque nano oscillators, which operate on space scales of nanometers and at gigahertz frequencies, and that aggregations of them may transmit and process digital information by means of spin waves.  The spin torque phenomena discussed here have been discovered and developed by physicists only recently (see references to this development in \cite{HKM2011}), and measurement of spin waves is difficult in practice, although measuring activity of an STNO is possible.  In any case, detection of possible spin waves in the brain seems far off.

On the other hand, the model here is identical to a canonical model of an AH bifurcation, complete with appearance of oscillations and exchanges of stability. This fact suggests that oscillator systems containing an AH bifurcation structure \cite{wcnn}, including those from neuroscience, may perform similar computations.  Since bursts of action potentials are known to carry amplitude, phase and frequency information (e.g., see \cite{desai}), it is possible to speculate that neural circuits may use action potentials for communicating and performing similar logic computations on the space and time scales of neural activity.

\section{Formal Logic and Binary Arithmetic.}
The basis of arithmetic in computers is performing addition of binary numbers bit by bit, while keeping track of carry-over. 
Binary addition may be accomplished using logic gates that are electronic embodiments of the basic operations from formal mathematical logic of OR, AND and NAND. These may be combined to perform all arithmetic operations using NAND logic. In the following, the symbols $a$ and $b$ may be used as generic logical entities; they may be binary digits, sets, signals, or represent other relevant concepts.

Disjunction is calculating $a$ OR $b$, which is written as
$
a \lor b.
$
Conjunction is calculating $a$ AND $b$, which is written as
$
a\land b.
$
NAND calculates the negation of $a$ AND $b$, which is written as
$
a\,\bar\land\, b \equiv \lnot(a\land b),
$
where $\lnot$ denotes negation. Note that in NAND logic, the logic function NOT may be calculated using the fact that $\lnot a \equiv a \,\bar\land\, a$.  All of these operations are defined by truth tables.

The sum of the two binary digits $a, b$ (i.e., 0 or 1), may be accomplished using the Exclusive OR logic function, XOR$(a,b)$, which may be expressed in NAND logic. This is discussed in Section 2.2.

\subsection{Binary Digits as Continuous Waveforms.}

The digit 1 is represented here by the signal 
\beq\label{eq:p}
p(\omega_p t) \equiv A_p\cos (2\pi\omega_p t + \phi_p)
\eeq
for some given amplitude $A_p$, frequency $\omega_p$, and phase deviation $\phi_p$; and, the digit 0 is represented by the signal 
$
n(\omega_p t)\equiv A_p\cos(2\pi\omega_p t\pm\pi+ \phi_p).
$
von Neumann \cite{JvN} showed how logic statements may be calculated using various superpositions of these signals along with the nonlinear circuitry. For the remainder we take $\phi_p=0$ without loss of generality. Also, other wave forms for $p$ may be used, but this is not explored further here.

Encoding a binary digit $a$ in terms of phase variables is done using the formulas
$$
A= -\cos a\pi \textrm{  and  } a=1-\arccos(A)/\pi,
$$ 
where $A=+1$ if $a=1$ and $A=-1$ if $a=0$. Then, for example,
if $A=1$, $AA_p\cos(\omega_p t) = p(\omega_p t)$, which represents the binary digit 1,  and if $A=-1$, $AA_p\cos(\omega_p t) = n(\omega_p t)$, which represents the binary digit 0.

We write L$(a,b)$ to represent a particular logic function operating on $(a,b)$. In particular,
$$
\begin{array}{cccc}
a \lor b:& \textrm{L}_{\lor}(a,b) &=& \textrm{sign}(1 + A + B)\\
a \land b:& \textrm{L}_{\land}(a,b) &=& \textrm{sign}(-1+ A + B)\\
a \,\bar\land\, b:& \textrm{L}_{\bar\land}(a,b) &=& \textrm{sign}(1-2A-2B). 
\end{array}
$$
Since there is an odd number of terms in each case, these functions have the value $\pm 1$. These (nonlinear) functions produce the correct truth tables for  OR, AND, and NAND:
$$
\begin{array}{cccccccccc}
a&b&a\lor b&a\land b&a \bar\land b&A&B&L_{\lor}&L_{\land}&L_{\bar\land}\\
0&0&0&0&1&-1&-1&-1&-1&+1\\
1&0&1&0&1&+1&-1&+1&-1&+1\\
0&1&1&0&1&-1&+1&+1&-1&+1\\
1&1&1&1&0&+1&+1&+1&+1&-1
\end{array}
$$

\subsection{Computation of Logic Functions.}

\textit{Disjunction} involves combining two inputs $\delta_1(t), \delta_2(t)$, where the possible inputs are $\delta_j(t) = p(\omega_p t)\textrm{ or } n(\omega_p t)$ for $j=1,2$, with a reference signal $p(\omega_p  t)$. The function
$$
p(\omega_p t)L_{\lor}(A,B)
$$
gives the equivalent of $\delta_1(t)\lor\delta_1(t)$. The output is proportional to $p$ if and only if at least one of the input signals (not counting the reference signal) is $p$; that is, if and only if at least one of the input digits is 1, not counting the reference signal. 

\textit{Conjunction} is similar. It combines two inputs $\delta_1(t), \delta_2(t)$, but now with reference signal $n$ as
$$
p(\omega_p t)L_{\land}(A,B),
$$
which gives $\delta_1(t)\land\delta_2(t)$.
The output is proportional to $p$ if and only if both the input signals are $p$; that is, if and only if both the input digits are 1.

\textit{NAND} combines four inputs with a reference signal $p(\omega_p  t)$:
$$
p(\omega_p t)L_{\bar\land}(A,B),
$$
which gives $\delta_1(t)\bar\land\delta_2(t)$.
The output is proportional to $p$ if and only if at least one of the input signals is $n$; that is, if and only if at least one of the input digits is 0.

\section{Spin Torque Nano Oscillators.}

A spin torque nano oscillator is created by passing a polarized electrical current through a contact on a ferromagnetic thin film. The film may be a few nanometers (nm) thick and the contact size may be in the range of 10 -100 nm. The Landau-Lifshitz-Gilbert-Slonczewski (LLGS) system of equations describes the magnetization dynamics in the film. Under natural conditions shown in \cite{HKM2011}, this system may be reduced to a single nonlinear Schr\"odinger equation for a complex function $u(x,y,t)$ where the film is described by spatial (in-plane) variables $(x,y)$, $t$ is time in some appropriate units (e.g., pico-seconds), and the components of the magnetization vector projected to the plane of the film are described by the vector $(\Re u, \Im u)$. The model is
\beq\label{eq:hLLGS}
\imath\,\frac{\pa u}{\pa t} = D\nabla^2 u + \omega u + \imath\,\left(\lambda +C(x,y,t) - b |u|^2\right)u
\eeq
where $\imath^2=-1$ and $u(x,y,t)$ represents the magnetization at a point $(x,y)$ in the film. This model is derived and studied with appropriate boundary and initial conditions  in \cite{HKM2011,MHK2014}. In the units used there, appropriate data are 
$$
 D = 1 + 0.01\imath, \quad\omega = 0.15, \quad\lambda= -0.1,  \quad b =0.1.
 $$
The parameter $\lambda$ accounts for damping of elements in the film and for spin torque, and the wave term $D\nabla^2 u$ describes torque exchange. The forcing term  $C(x,y,t)$ accounts for external modulation of an applied electrical current. Spatially propagating solutions of (\ref{eq:hLLGS}) are referred to as being spin waves.

When attention is restricted to a set of $N$ isolated STNO that are sufficiently separated that the wave term  may be ignored ($D = 0$), the nonlinear Schr\"odinger equation (\ref{eq:hLLGS}) may be approximated by a system of  ordinary differential equations for the magnetization at each STNO. In that case, the magnetization at the $j^{th}$ STNO is represented by $u_j$, for $j=1,\dots,N$, and the model (\ref{eq:hLLGS}) reduces to a system of nonlinear ordinary differential equations
\beq\label{eq:dhLLGS}
\imath\,\frac{du_j}{dt}= \omega u_j + \imath\,(\lambda  - b |u_j|^2)u_j + \imath\,C_{j}(t) u_j
\eeq
for $j=1,\dots,N$. The vector $C=(C_j)$ describes external input of the various STNO in the system.

The designs presented here reveal how to choose $C$ to perform particular specific tasks. 
A good test of computational capabilities of STNO arrays is in calculating elementary logic statements that may be used for computation in binary arithmetic.

\subsection{Embodiment of Binary Calculation using STNO.}

\begin{figure}[htb!]
\begin{subfigure}{.5\textwidth}
 \centerline{\psfig{figure=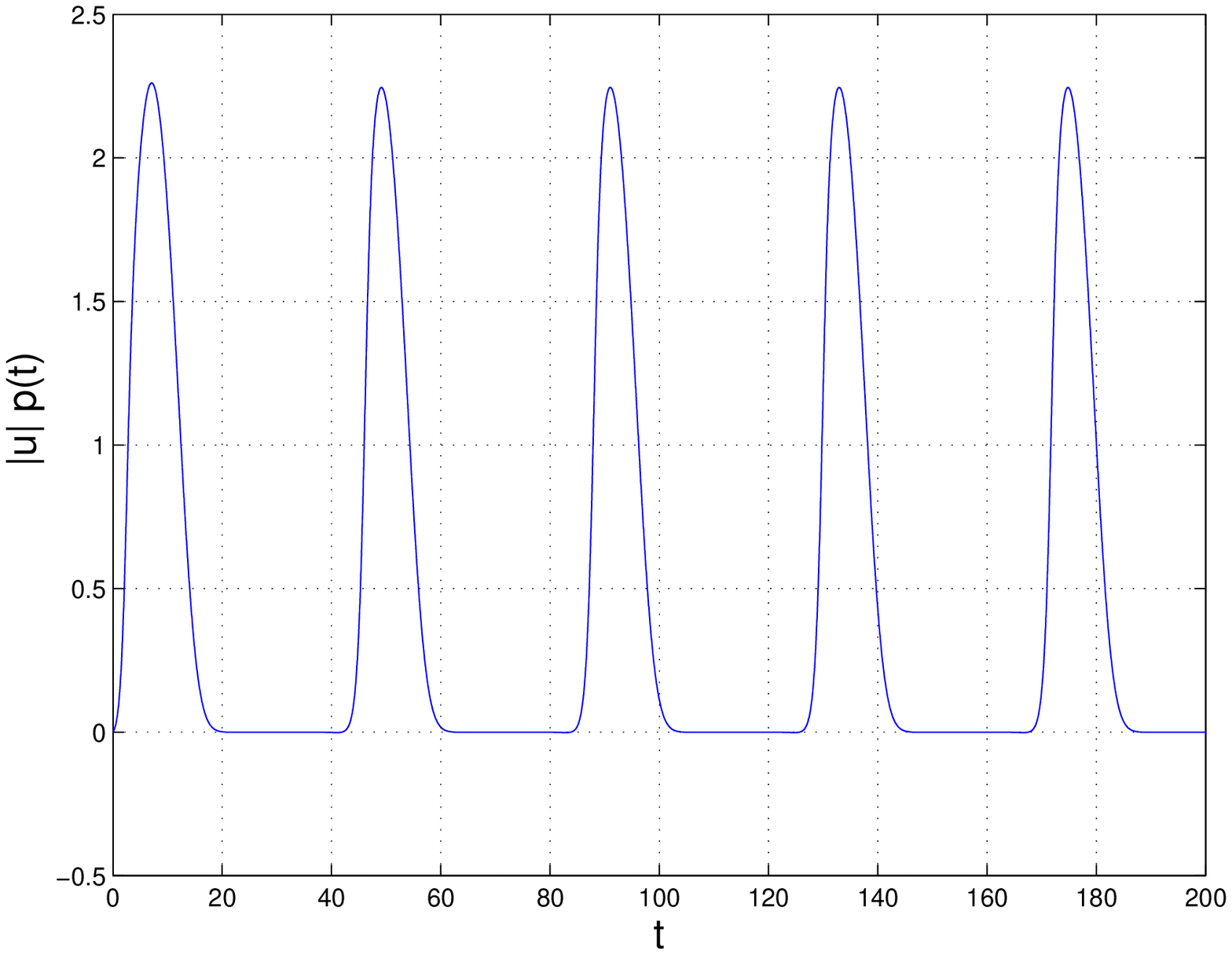,width=2in}}
  \caption{$C(t)\propto p(\omega_p\,t)$}
\end{subfigure}%
\begin{subfigure}{.5\textwidth}
  \centerline{\psfig{figure=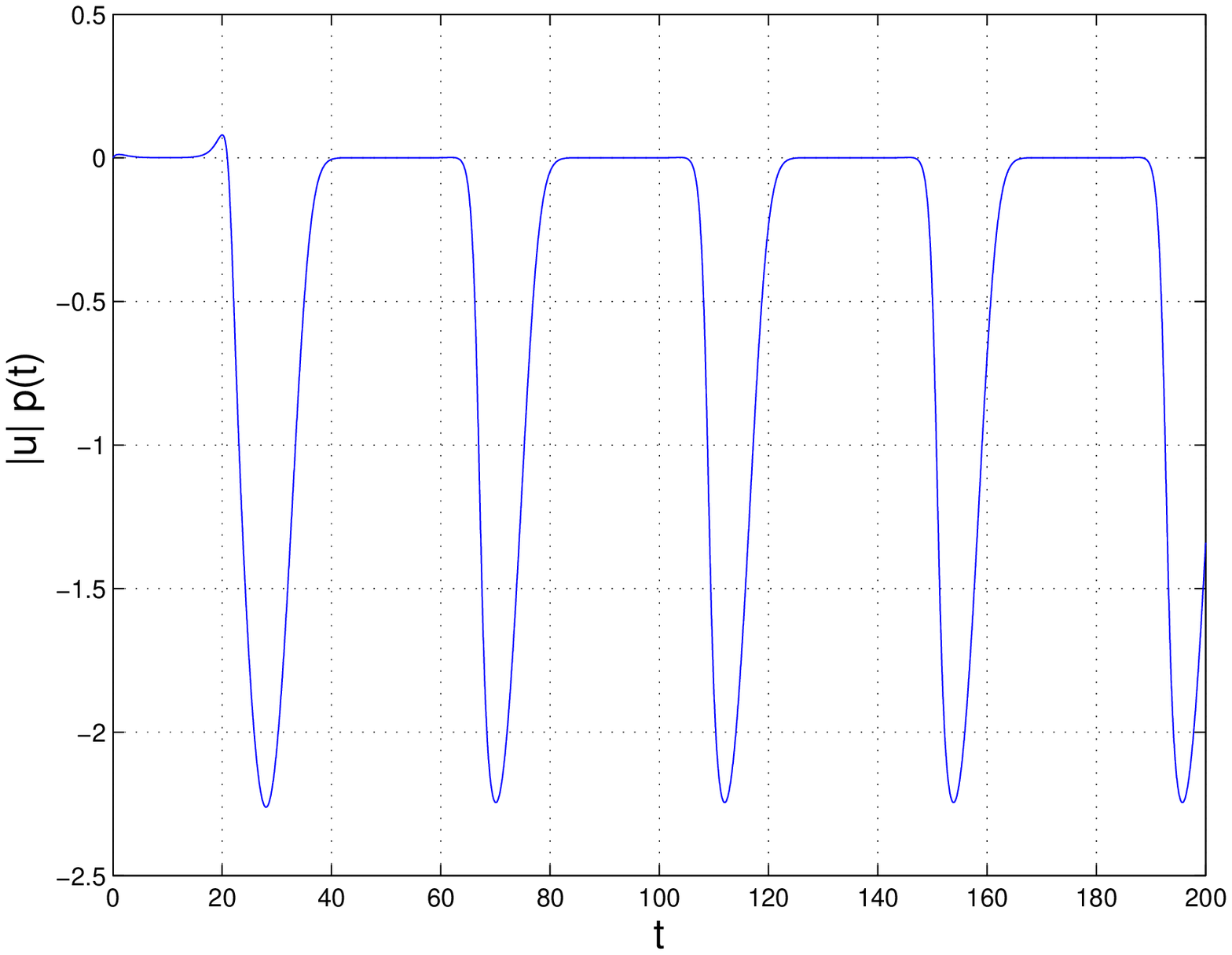,width=2in}}
  \caption{$C(t)\propto n(\omega_p\,t)$}
\end{subfigure}
\caption{Solving (\ref{eq:dhLLGS}) with $C(t)\propto p(\omega_p\,t)$ and plotting $|u|p(\omega_p\,t)$ (left panel), and solving (\ref{eq:dhLLGS}) with $C(t)\propto -p(\omega_p\,t)$ and plotting $|u|p(\omega_p\,t)$ (right panel).}
\label{fg:majSen140412}
\end{figure}

Equation (\ref{eq:dhLLGS}) may be analyzed for stability by observing that in polar coordinates the amplitude $r_j= |u_j|$ satisfies a (scalar) Riccati equation; details are not shown here. We simply observe that $r_j^2$ tries to track $(\lambda+C_j(t))/b$, so we expect that the solutions in the two cases where $C\propto p$ or $C\propto n$ will be in anti-phase. Note that $r_j^2\to 0$ when $\lambda+C_j(t)\le 0$; that is, there is an exchange of stabilities as $C_j(t)$ passes through the Andronov-Hopf bifurcation at  $C_j(t) = -\lambda$. These behaviors are apparent in the simulations in Figure~ \ref{fg:majSen140412}.

von Neumann \cite{JvN} showed how to use oscillators based on nonlinear inductance and nonlinear capacitance, such as crystal diode circuits, to evaluate logic functions. While STNO were not known to him, they are similar in mathematical terms to the nonlinear oscillators he mentions.

Revealed next is how a single STNO may embody his invention. Consider the model (\ref{eq:dhLLGS}) 
with the applied forcing having the form 
$$
C(t)=p(\omega_p t)\textrm{L}(A,B)
$$ 
for a logic function L, as described earlier. The output from such a system is a complex signal $u(t)$, and the logic result may be determined from the correlation
\beq\label{eq:up}
|u(t)| p(\omega_p t).
\eeq
This is illustrated by the computer simulations in Figure~\ref{fg:majSen140412}. The spikes of the correlation will be negative when the outcome is the digit 0 and positive when the outcome is the digit 1. 

\subsection{Computation of Iterated Logic Functions.}

Although von Neumann did not specify how oscillators might be connected to perform particular (complicated) logic calculations, the following shows one way to do this. 

In general, each of $a\lor b, a\land b$ and $a\,\bar\land\, b$ may be represented by a component of $u$ in (\ref{eq:dhLLGS}), perhaps with input from others as needed in building a logic statement. In such cases, the vector $C$ is augmented to describe the connection stencil and to tell how information is to be passed from one calculation to the next. We modify (\ref{eq:dhLLGS}) by introducing auxiliary variables ($v_j$) to represent outputs:
\begin{eqnarray}
\imath\,\frac{du_j}{dt}&=& \omega u_j + \imath\,(\lambda  - b |u_j|^2)u_j + \imath\,C_{j}(t,v) u_j\\
\tau_j\frac{dv_j}{dt}&=&-v_j+\textrm{sign}(p(\omega_p t)|u_j(t)|)\nonumber
\end{eqnarray}
for $j=1,\dots,N$, where $\tau_j$ is a time constant (possibly $\tau_j=0$). The variables $v=(v_1,\dots,v_N)$ describe filtered outputs of the various STNO, and $v_j$ represents the correlation of $|u_j|$ with $p(\omega_p t)$. Here we take components of $C$ in the form
$$
C_j(t,v)=p(\omega_p t)\textrm{L}(\textrm{sign}(v_k(t)),\textrm{sign}(v_l(t)))
$$
for some pair of inputs from computations at sites $k$ and $l$.  For example, to calculate 
$$
\textrm{XOR}(a,b)=(a\,\bar\land\,(a\,\bar\land\, b))\,\bar\land\,(b\,\bar\land\,(a\,\bar\land\, b))=(a\land\lnot b)\lor(\lnot a \land b),
$$
we may take
$$
C_1=p(\omega_p t)\textrm{L}_{\land}(-A,B), C_2 = p(\omega_p t)\textrm{L}_{\land}(A,-B), C_3 = p(\omega_p t)\textrm{L}_{\lor}(\textrm{sign}(v_1),\textrm{sign}(v_2)).
$$
The output $\textrm{sign}(v_3)$ gives the value of XOR($a,b$).

\subsection{Multiplex Computation by a Single STNO}
\begin{figure}[htb!]
\begin{subfigure}{.5\textwidth}
 \centerline{\psfig{figure=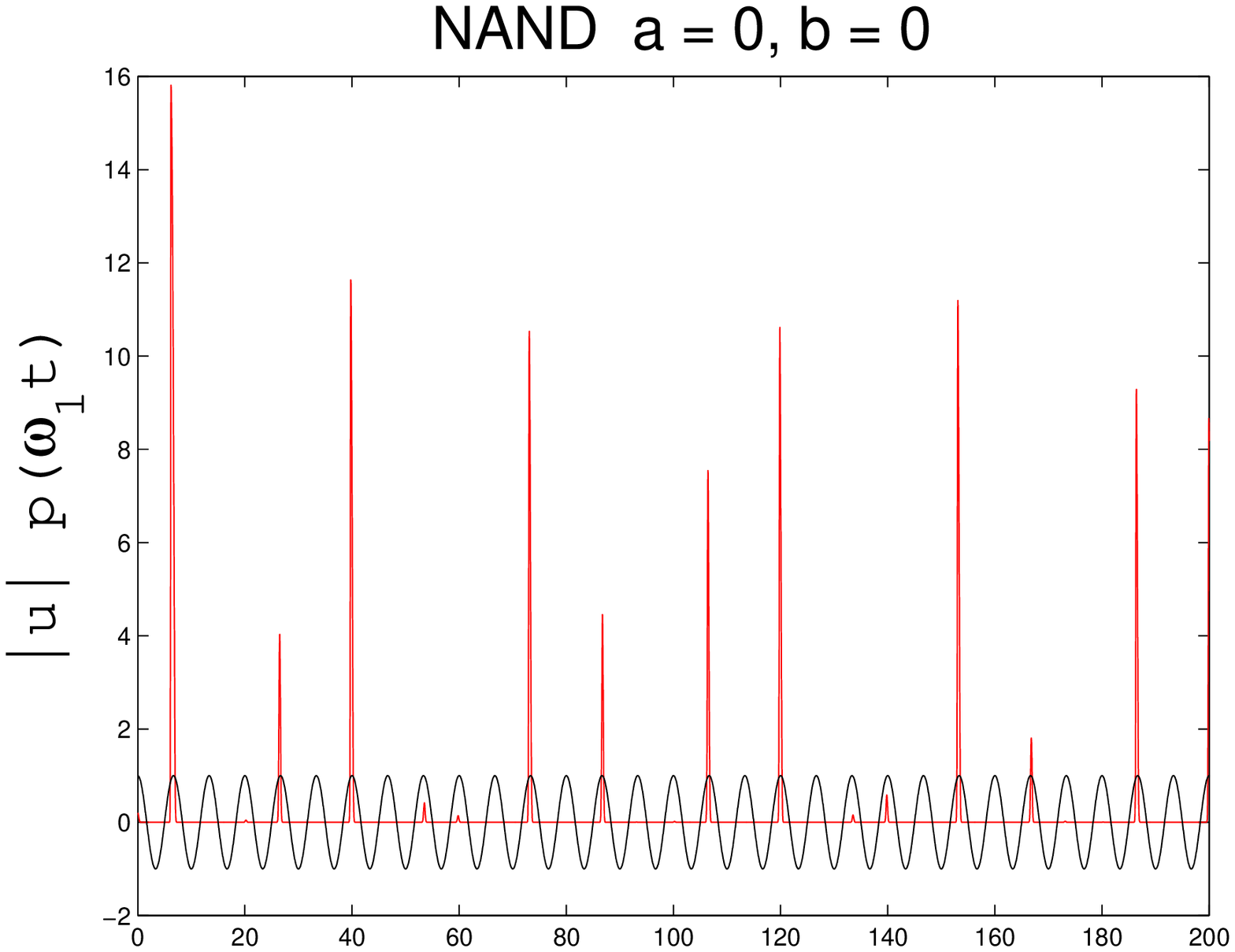,width=2in}}
  \label{fig:sub1}
\end{subfigure}%
\begin{subfigure}{.5\textwidth}
  \centerline{\psfig{figure=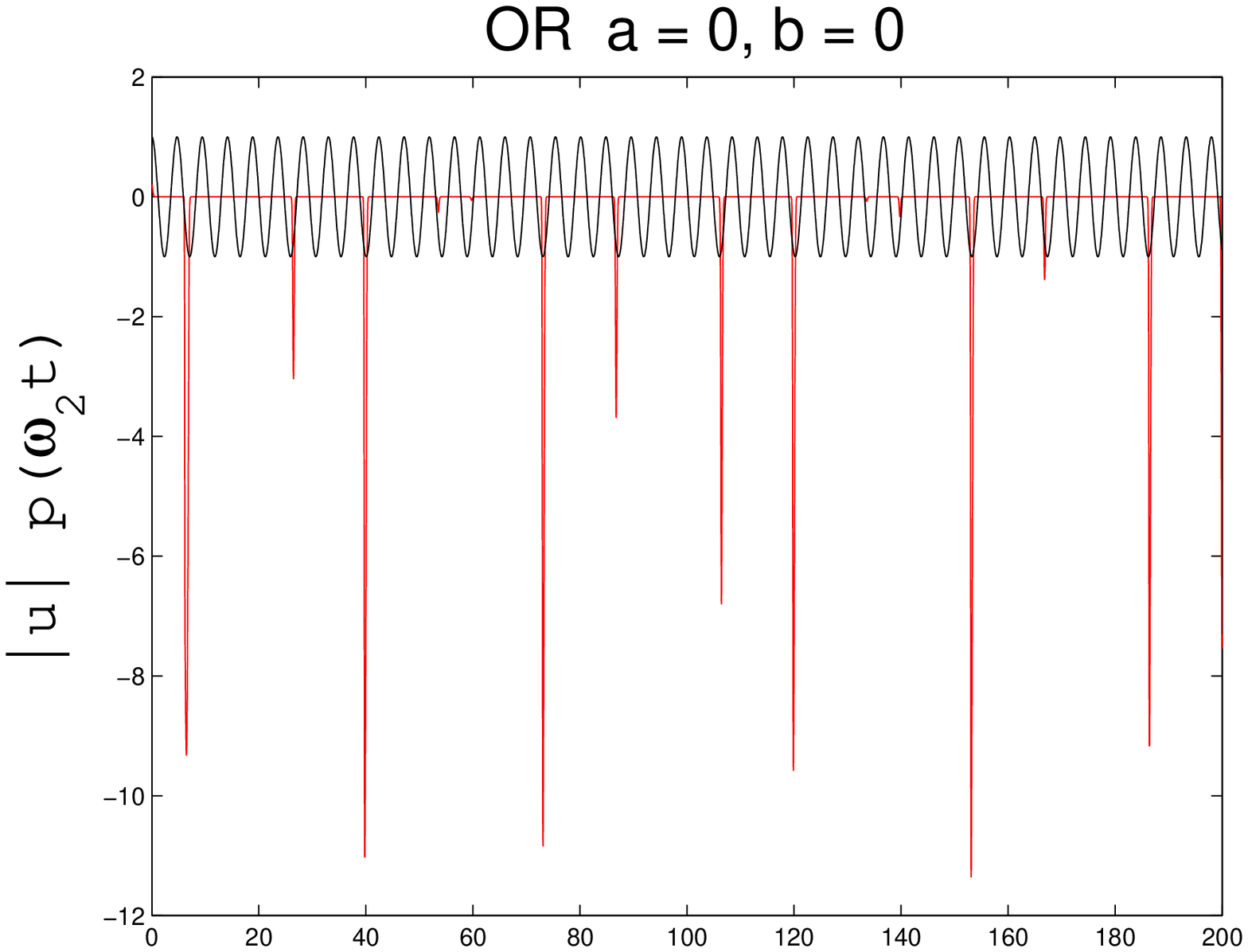,width=2in}}
  \label{fig:sub2}
\end{subfigure}
\caption{Solving (\ref{eq:dhLLGS}) for $u$ with L$_1(a,b)=$ L$_{\bar\land}(a, b)$ and L$_2(a,b)=$ L$_{\lor}(a, b)$  and plotting $|u(t)|p(\omega_1t)$  (left panel) and plotting $|u(t)|p(\omega_2 t)$  (right panel). The results are that $0\bar\land 0 =1$ (left) and $0\lor 0 = 0$ (right).}
\label{fg:MUX}
\end{figure}

The ideas of the preceding sections are extended here to design a device that can perform multiple logic computations simultaneously. A single STNO ($N=1$ in equation (\ref{eq:dhLLGS})) may be used to simultaneously evaluate two or more logical functions by using frequency multiplexing. Consider the forcing function
$$
C(t) =\cos(2\pi\omega_1\, t)\,\textrm{L}_1(A,B) + \cos(2\pi\omega_2\,t)\,\textrm{L}_2(A,B)
$$
where the channel frequencies $\omega_1$ and  $\omega_2,$ are sufficiently separated (in the simulations, we use $\omega_2/\omega_1 = \sqrt{2}$), and the functions L$_1, $ L$_2$ describe whatever logic functions are to be computed. The examples in Figures~\ref{fg:MUX} show the circuit performing NAND and OR at the same time.

In Figure~\ref{fg:MUX} for NAND and OR, respectively, the 0 and 1 output spikes for their respective logic functions occur at times that are determined by resonance conditions between $\omega_1$ and $\omega_2$. That is, the system waits until there is an appropriate match between extrema of $p(\omega_1\, t)$ and $p(\omega_2 t)$, and the results are spikes at those times. 

\section{Information Processing by Spin Waves.}
Spin waves can selectively transmit the outcome of a logic computation from one site to another with there being no electrical connection between oscillator elements. Rather connections are established by spin waves.

The wave term ($D\nabla^2u$) in (\ref{eq:hLLGS}) becomes important when some STNO are near each other. In particular, it is known that two nearby STNO (within the order of a wavelength) will phase lock, the connection being through spin waves, and the pair may generate a complicated interference pattern of spin waves \cite{MHK2014}. We consider one example of this here.

Four pairs of STNO are placed on a film as shown in Figure~\ref{fg:placement}. Pair 1,2 is forced with $p(\omega_p t)$ and pair 3,4 is forced anti-phase with $n(\omega_p T)=-p(\omega_p t)$. The responses of the other four (5-8) are not forced other than by spin waves. That is, the pairs 1,2 and 3,4 are taken to be sources of signals, and the pairs 5,6 and 7,8 are taken to be detectors. 

We define
$$
C(t,x,y)\propto p(\omega_p\,t)(\chi_1(x,y)+\chi_2(x,y)-\chi_3(x,y)-\chi_4(x,y))
$$
where $\chi_j(x,y)$ is the indicator function of the contact at site $j$ (i.e., it is one for $(x,y)$ in the contact, but otherwise zero) for $j=1,2,3,4$.  The placement of the STNO is shown in Figure~\ref{fg:placement}.
\bef[htb!]
\centerline{\psfig{figure=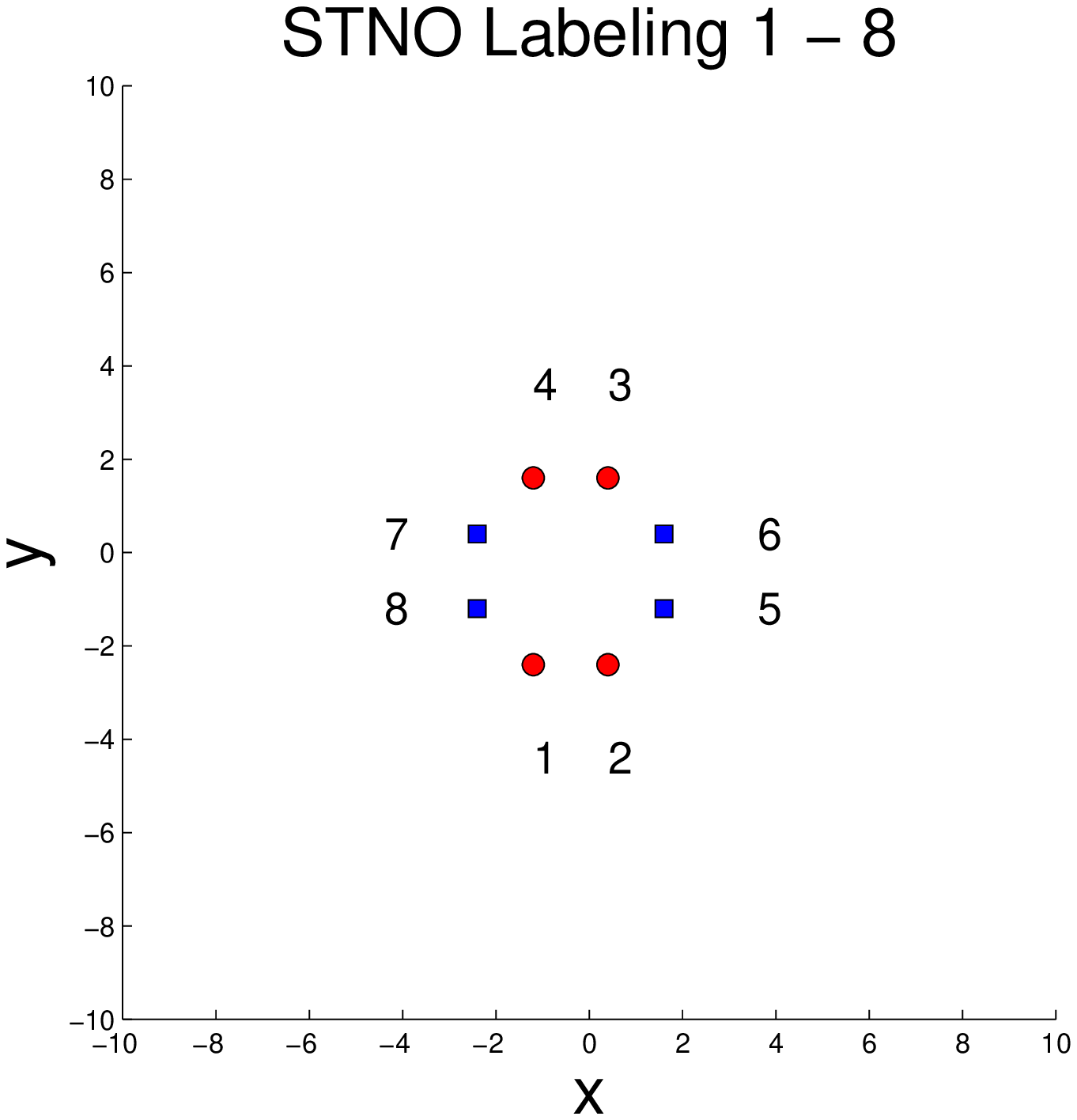,width=2in}}
\caption{STNO placement labels in the following simulation are: Bottom: pair 1,2 (forced by $p$). Top: pair 3,4 (forced by $n$). Pairs 1,2 and 3,4 are referred to as being sources. Right: pair 5,6 (not forced other than by spin waves). Left: pair 7,8 (not forced other than by spin waves). Pairs 5,6 and 7,8 are referred to as being detectors. Dots denote approximately 10 nm.}%
\label{fg:placement}
\eef

\begin{figure}[htb!]
\centerline{\psfig{figure=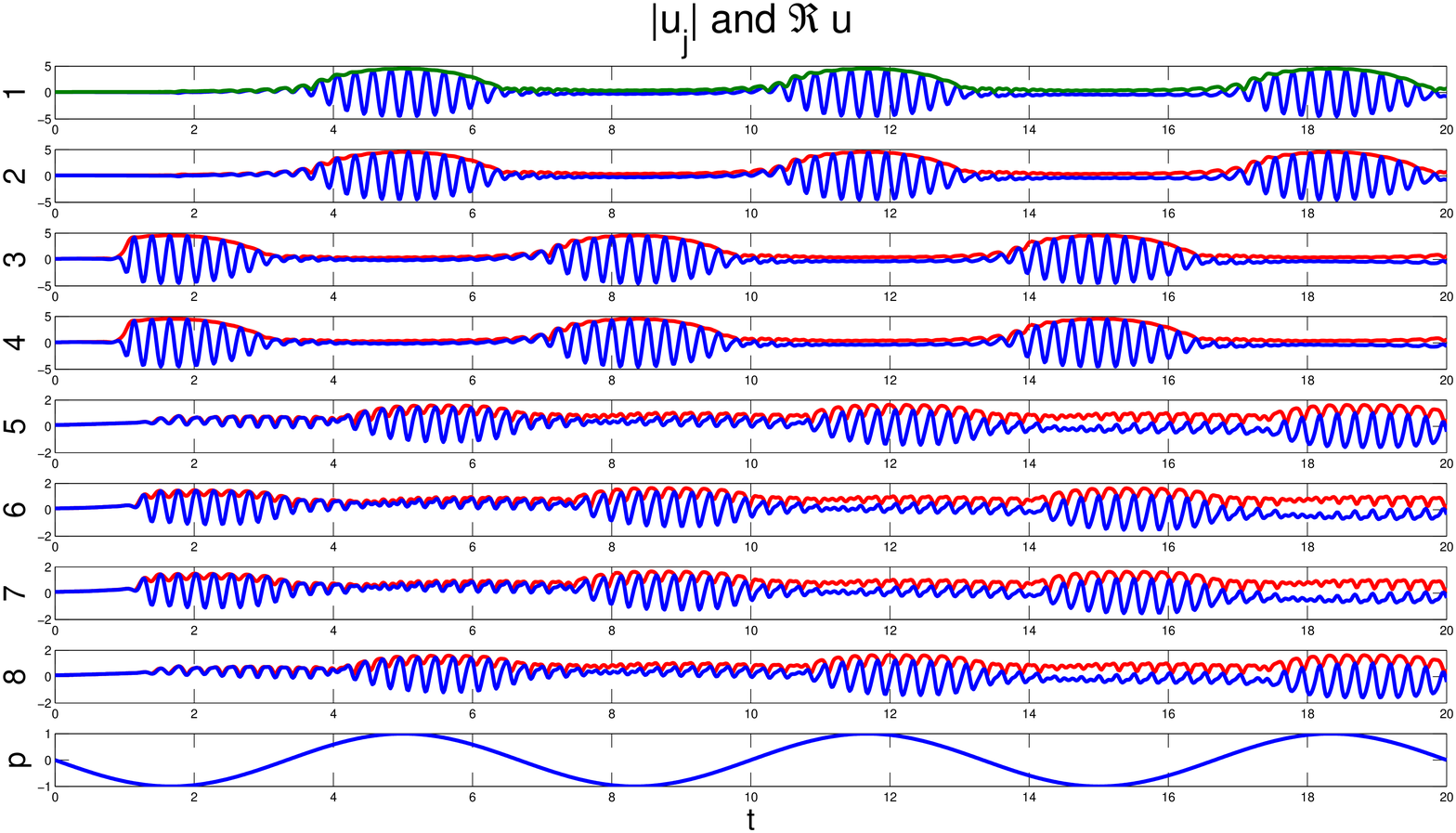,width=3in}}
\caption{Solving (\ref{eq:hLLGS})
and plotting the envelope of oscillation ($|u|$) and $\Re u$ in each case. Detector sites 5-8 are not forced, but receive input from spin waves. Sites 1,2 are forced and 3,4 are forced anti-phase to 1,2. Sites 5,8 are nearest to 1,2, and sites 6,7 are nearest 3,4.  These lock to their nearest forced neighbor. For example, the forced bursts in 1,2 correspond to the digit 1 and those in 3,4 correspond to the digit 0. They entrain the detector STNO so that 5,8 produce the digit 1 and 6,7 the digit 0.  The selective forcing by spin waves is due to the radiation pattern created by a forced pair in this configuration. Note the slight delay in firing of the detector sites, which is due to propagation time of spin waves from their source. This simulation shows that the phase and frequency of bursts are communicated to detectors by spin waves.}%
\label{fg:desai}
\end{figure}

The simulation in Figure~\ref{fg:desai} shows that bursts appear in the forced STNO and that the nearest of the detector STNO phase lock to the forced ones, The simulation illustrates that the phase of the bursts may carry information. Logic values may be determined as earlier by correlation with the forcing function $p$. Coupling from sources to detectors is through spin waves, not through electrical contacts, in this simulation. We consider elsewhere development of other STNO arrays that are coupled through spin waves. 

\section{Discussion}
This paper presents elements needed to fully implement von Neumann's invention in a spin torque system. Presented here are several applications where his invention may be implemented and extended using STNO. For example, his patent does not give details of how oscillators might be connected, but the work in Section 2.2 gives one useful embodiment of electrical connections. In another example, he suggests connecting oscillators using wave guides or the like, while spin waves may be used here for connectivity. The examples indicate the scope of applications of the STNO paradigm, and they suggest numerous additional applications and extensions thereof. 

An intriguing aspect of STNO arrays is that they do some neuron-like things, but at gigahertz frequencies and nanometer space scales, compared to millimeter space scales and frequencies typically to 100 hertz in a brain. Most results for oscillatory or weakly connected neural networks in the literature, for example \cite{wcnn, olga}, may be embodied in the STNO paradigm. These systems are quite stable to random and small systematic perturbations, partly due to the phenomenon of phase locking.

At the time of von Neumann's work, in the early 1950s, it was not known what possibilities for high performance computing the newly discovered transistor would provide, and he explored alternatives to vacuum tube circuits for computing using microwaves. We do not know the relation between his patent and his interest in the brain, but certainly he was thinking about computers and brains as described in his book \cite{jvnBrain}. 

Can neural networks perform logic computations using action potentials? It is known that neural networks perform switching and control functions,  as for sound location and central pattern generation. The results revealed here suggest another aspect of what physiological neurons might say to each other. Action potential bursts may contain amplitude, phase and frequency information  \cite{desai}, and it is plausible that neural networks having the AH bifurcation structure described here, such as the Hodgkin-Huxley model, may perform logic calculations using bursts or other action potential phenomena. There are many stable background oscillations in the brain that might serve as references for encoding and decoding information (e.g., play the role of $p$ in the correlations here). While it is also possible that magnetic oscillations and waves play a role in computation in the brain,  nano magnetic experiments in physical systems are difficult, and detecting these phenomena in biological systems is beyond the present state of the art.

\end{document}